\documentclass[a4paper,12pt]{article}
\usepackage{amsmath,amsfonts,amssymb,cite}

\begin{document}

\begin{center}
{\Large\bf Towards reconciling the holographic and lattice descriptions of radially excited hadrons}
\end{center}

\begin{center}
{\large S. S. Afonin\footnote{E-mail: \texttt{s.afonin@spbu.ru}}}
\end{center}

\begin{center}
{\it Saint Petersburg State University, 7/9 Universitetskaya nab.,
St.Petersburg, 199034, Russia}
\end{center}

\begin{abstract}
Within the framework of AdS/QCD models, the spectra of radially excited hadrons
are identified with towers of Kaluza-Klein (KK) states in a putative dual theory.
The infinite number of KK states is indispensable to provide correct high energy
asymptotics of correlation functions in QCD.
It is known, however, that the KK modes of dual theory must be qualitatively different
from real hadrons. And what is more important, the radially excited states appear in lattice
calculations not as "excitations" of some ground state, but rather as independent states
coupled to higher dimensional QCD operators --- the larger is a basis of interpolating operators,
the larger set of states can be resolved. A question arises whether it is possible to reconcile
the holographic and lattice descriptions of radially excited hadrons?
We propose a new phenomenological "consistency test" for bottom-up holographic models:
If the KK spectrum of massive
5D fields dual to higher dimensional operators in QCD coincides with the conventional
radial KK spectrum, then the holographic KK states can be identified with real excited mesons in the large-$N_c$
limit of QCD.
We demonstrate that the Soft Wall holographic model passes this test while
the Hard Wall model does not.
\end{abstract}

\newpage

\section{Introduction}

The confinement in QCD is known to lead to a rich spectrum of excited hadrons.
A complete theoretical understanding of this spectrum is still missing despite of
many fruitful ideas and models put forward in the last half of century. The problem
becomes especially acute in the case of hadrons composed from light quarks
where one observes plenty of higher spin and radial excitations. The corresponding
hadron resonances usually have large decay widths and this causes many difficulties
not only with experimental extraction of their characteristics but also with a clear
theoretical identification of the given objects. In this situation, it is very useful
to consider a limit where many hadron resonances become well-defined
particles while the underlying theory remains qualitatively similar. Quite remarkably,
such a limit indeed exists in QCD --- the large-$N_c$ (called also planar) limit~\cite{hoof},
in which the quark-antiquark states become stable particles: Both their strong decay
width and corrections to masses are suppressed by $1/N_c$. Baryons represent heavy
objects in this limit~\cite{wit} (their mass behaves as $\mathcal{O}(N_c)$ while the
mass of quark-antiquark mesons scales as $\mathcal{O}(1)$) and will be of no interest
for us. Matching the obtained meson theory to the perturbative QCD at small 't Hooft
constant $\lambda\doteq g^2N_c$ shows that the theory at small $\lambda$ must contain
an infinite number of states for each quantum number~\cite{wit} --- the "radial" excitations
in the language of non-relativistic potential models.

It has long been suspected that planar QCD is equivalent to some string theory.
This idea got a new push with the advent of AdS/CFT (or more generically holographic)
correspondence~\cite{mald,witten,gub}. The use of holographic conjectures has now
penetrated many branches of theoretical physics including the real QCD
(this branch of research is referred to as
holographic QCD or AdS/QCD). Within the holographic QCD, the infinite number of mesons
with identical quantum numbers expected in the large-$N_c$ limit is modelled by an
infinite tower of Kaluza-Klein (KK) excitations. The most popular phenomenological
models of this sort are the Hard Wall (HW)~\cite{son1,pom} and Soft Wall (SW)~\cite{son2}
holographic models. They have enjoyed an unexpected phenomenological success ---
the HW model was particularly successful in description of physics related
to the chiral symmetry breaking in QCD and of hadron form factors while the SW one turned out to be very convenient
in description of phenomenology of string-like linear Regge trajectories.

The modeling of radially excited mesons by KK modes looks, however, too simplistic ---
the former are highly complicated dynamical objects in QCD while the latter are rather
simple states arising from extra dimension. The dramatic difference between the KK-like
and QCD-like states is discussed in detail in Ref.~\cite{csaki}. The main point consists
in observation that the KK modes are deeply bound states sensitive to short distance
interactions and at collisions producing events with mostly spherical shapes while mesons
are extended states sensitive to large distance interactions and producing
characteristic jets. The underlying reason is that the latter are defined at small
't Hooft coupling $\lambda$ while the former at large $\lambda$ where the existence
of holographic duality can be motivated. The theories at small and large $\lambda$
turn out to be qualitatively different. In section~5, we will add to this discussion
by indicating rather hard difficulties with a correct definition of propagating
KK modes (except the lightest one) because the bulk space in the holographic duality
is not exactly the AdS one.

We see thus that the straightforward identification of KK states with the real
large-$N_c$ mesons --- what is always implied in AdS/QCD models --- has problems.
To address these problems, it is instructive to recall how the radially excited states
are described from first principles. Within the lattice calculations, these
states appear not as "excitations" of some ground state, but rather as independent states arising
from Higher Dimensional Operators (HDO) in QCD in the sense that the larger basis of interpolating operators
is included into simulations, the larger number of states can be resolved~\cite{Dudek}.
For a consistent holographic description of real mesons, it looks therefore natural
to introduce massive fields dual to HDO in QCD. These fields will
have their own discrete spectra of normalizable modes. Our idea is the following:
If the KK spectrum of each of new heavier fields coincide with the original KK spectrum since
some mode of original KK spectrum,  we can draw a one-to-one correspondence between each higher KK mode
and a set of QCD operators leading to the mass of this mode. We get then a self-consistent physical picture ---
on the one hand, the infinite tower of formal KK states provides a correct high energy asymptotics
for hadronic correlation functions (the importance of this requirement for internal consistency
of the model was recently emphasized in Refs.~\cite{MMM}), on the other hand, each KK state in this
tower can be traded for a state interpolated by a definite set of QCD operators and thus interpreted
as a real excited meson in the large-$N_c$ limit.

We will demonstrate how the outlined scheme works in the SW model.
In the case of HW model, however, the spectra of KK states and states emerging from HDO are different.
We interpret this as another one argument (in addition to
non-Regge like behavior and absence of power-like corrections in the OPE) that the KK
states of HW model cannot be identified with real hadrons. It is interesting to observe,
however, that the spectrum of ground states related with HDO
rises much slower with excitation number than the spectrum of radial KK states and qualitatively
approaches the rise of experimental radial trajectories.

Finally it should be mentioned that inclusion of new 5D fields dual to higher dimensional operators 
in 4D gauge theory improves the consistency of AdS/QCD models from the viewpoint of general principles
of gauge/gravity duality. We remind the reader that within the original AdS/CFT correspondence
conjectured in the type~IIB superstring theory, 
for description of 4D gauge theory one needs not only low-energy supergravity on AdS$_5$
but also the whole infinite tower of massive KK states on AdS$_5\times$S$_5$ which appear after
the KK-compactification on 5D sphere~\cite{witten}. These 5D KK states are "enumerated" by
scaling dimensions of higher-dimensional operators in 4D gauge theory. The given point
is of crucial importance in establishing a complete mapping of physical degrees of freedom between
the two dual theories.

The paper is organized as follows. In section~2, we begin with a very brief review on
the HDO in QCD. Then we apply our idea to
the SW and HW holographic models in sections~3 and~4 correspondingly.
Further discussions are contained in section~5 and we conclude in section~6.

\section{Higher dimensional QCD operators}

The general idea that radially excited hadrons should arise from
coupling to HDO in QCD is rather old and has appeared from time to time in the literature.
It was used, for instance, for construction of some extended low-energy
quark models~\cite{qqm,chizh}. And what is the most important, this idea is explicitly realized
in the lattice calculations of radial excitations~\cite{Dudek}.

Consider the following two basic quark currents,
\begin{equation}
\label{h1}
V_\mu=\bar{q}\gamma_\mu q=\bar{q}_\text{L}\gamma_\mu q_\text{L}+\bar{q}_\text{R}\gamma_\mu q_\text{R},
\end{equation}
\begin{equation}
\label{h2}
S=\bar{q}q=\bar{q}_\text{L}q_\text{R}+\bar{q}_\text{R}q_\text{L}.
\end{equation}
Here the Dirac spinor $q=q_\text{L}+q_\text{R}$ stays for $u$ or $d$ quark fields
and $q_\text{L,R}=\frac{1\mp\gamma_5}{2}q$. The isospin and $\gamma_5$ matrices
can be also inserted in~\eqref{h1} and~\eqref{h2} but this is not essential for
our discussions and will be dropped (but implicitly assumed where necessary).
What is essential is the different chiral structure of twist-2 vector current~\eqref{h1}
and twist-3 scalar current~\eqref{h2} --- they transform differently under
$SU_\text{L}(2)\times SU_\text{R}(2)$ chiral transformations. The chiral transformation
properties of QCD currents were discussed in detail in many papers and we refer to
the relevant literature (see, e.g.,~\cite{multiplets,shif}).

The twist-2 vector current~\eqref{h1} has been traditionally used for interpolation
of $\rho$ and $\omega$ mesons in QCD sum rules, lattice QCD and low-energy effective
field theories. Consider now the lightest $\rho_3$ excitation --- the resonance
$\rho_3(1690)$~\cite{pdg}. The leading twist spin-3 quark operator can be easily constructed
by insertion of covariant derivatives (the appropriate symmetrization is implied),
\begin{equation}
\label{h3}
V_{\mu_1\mu_2\mu_3}=\bar{q}\gamma_{\mu_1}D_{\mu_2}D_{\mu_3}q.
\end{equation}
The tensor current~\eqref{h3} is not conserved (as the scalar one~\eqref{h2} and many others) but the
experience of spectral QCD sum rules shows that the conservation is of no importance for finding the
relevant pole~\cite{svz}. The interpolating operator~\eqref{h3} repeats the chiral properties of~\eqref{h1}.
We can contract the last two Lorentz indices in~\eqref{h3} and get the twist-4 vector current
\begin{equation}
\label{h4}
V_\mu'=\bar{q}\gamma_\mu D^2 q.
\end{equation}
It is natural to expect that the current~\eqref{h4} couples to a spin-1 state lying in a mass range
close to $\rho_3(1690)$. Such a state does exists --- the resonance $\rho(1700)$~\cite{pdg}. There are many
other cases of approximate degeneracy between spin and radial excitations and this gives rise to a remarkable
general picture of degeneracies in the experimental spectrum of light non-strange mesons up to 2.5~GeV~\cite{shif,deg}.
We obtain thus a kind of "experimental" confirmation for the conjecture that the radially excited
states couple to HDO in QCD.

On the other hand, a vector interpolating current can be constructed also by insertion of covariant
derivative to the scalar current~\eqref{h2},
\begin{equation}
\label{h5}
\tilde{V}_\mu=\bar{q}D_\mu q=\bar{q}_\text{L}D_\mu q_\text{R}+\bar{q}_\text{R}D_\mu q_\text{L}.
\end{equation}
This operator inherits the chiral properties of the current~\eqref{h2} and its twist. In addition,
the currents~\eqref{h1} and~\eqref{h5} look different from the point of view of Lorentz group
since~\eqref{h5} can be represented on shell as~\cite{shif}
\begin{equation}
\label{h6}
\bar{q}D_\mu q \propto -\partial^\nu H_{\mu\nu},\qquad H_{\mu\nu}=\bar{q}\sigma_{\mu\nu} q,
\end{equation}
where $\sigma_{\mu\nu}=(\gamma_\mu\gamma_\nu-\gamma_\nu\gamma_\mu)/2i$. The antisymmetric tensor
current $H_{\mu\nu}$ transforms as $(1,0)+(0,1)$ while~\eqref{h1} has the Lorentz structure
$\left(\frac12,\frac12\right)$~\cite{chizh,multiplets,shif}. One can notice that the vector current $\partial^\nu H_{\mu\nu}$
is trivially conserved but this conservation is topological, i.e. of different nature than the conservation
of Noether current~\eqref{h1}. What $\rho$-meson does the operator~\eqref{h5} interpolate?
Since it has one covariant derivative, one can expect the corresponding state to lie between
$\rho(770)$ and $\rho(1700)$. There exists one well-established $\rho$-meson in this range ---
the resonance $\rho(1450)$~\cite{pdg}. One may expect also that the experimental study of this resonance
is more difficult because its production should be suppressed in $e^+e^-$-annihilation.
The Particle Data~\cite{pdg} indeed makes caution that $\rho(1450)$ is the name for a broad resonance
region rather than a definite resonance. The first suggestion to associate~\eqref{h6} with $\rho(1450)$
appeared within the extended Nambu--Jona-Lasinio model of Ref.~\cite{chizh}.

The higher spin Regge recurrences arise thus from two kinds of composite spin-$J$ operators
stemming from~\eqref{h1} and~\eqref{h2},
\begin{equation}
\label{h7}
V_{\mu_1\mu_2\mu_3\dots\mu_J}=\bar{q}\gamma_{\mu_1}D_{\mu_2}D_{\mu_3}\dots D_{\mu_J} q,
\end{equation}
\begin{equation}
\label{h8}
\tilde{V}_{\mu_1\mu_2\mu_3\dots\mu_J}=\bar{q}D_{\mu_1}D_{\mu_2}D_{\mu_3}\dots D_{\mu_J} q.
\end{equation}
The operators~\eqref{h7} interpolate the spin-$J$ quark-antiquark states in the Lorentz
representation $\left(\frac{J}{2},\frac{J}{2}\right)$ and have the chiral transformation
properties of usual vector current~\eqref{h1} (and for this reason emerge naturally in QCD
analysis of deep inelastic scattering via OPE), while the chiral and Lorentz properties
of~\eqref{h8} are different. Contracting $n$ times the Lorentz indices we get an interpolating
operator for the $n$-th radial excitation of corresponding spin-$(J-2n)$ meson.

For high enough canonical dimensions, several covariant derivatives in~\eqref{h7} and~\eqref{h8}
can be replaced by insertion of gluon field strength $G_{\mu\nu}$, the corresponding operators
will interpolate hybrid states. For instance, the operator
$\bar{q}\gamma^{\mu}G_{\mu\nu}D^{\nu}q$ couples to a scalar hybrid with the chiral properties
of vector current~\eqref{h1}. One can of course construct purely gluonic operators which are
chiral singlets. The leading twist-2 operators of this sort have the structure
\begin{equation}
\label{h9}
\tilde{G}_{\mu_1\mu_2\mu_3\dots\mu_J}=G^\rho_{\mu_1}D_{\mu_2}D_{\mu_3}\dots D_{\mu_{J-1}}G_{\mu_J\rho}.
\end{equation}

In relating holographic predictions to observable states one should keep in mind that the spectrum of
holographic models by itself does not know about the chiral and Lorentz structure of underlying
QCD operators, only canonical dimension and spin are essential.

\section{The Soft Wall model}

For demonstration of our main idea we will
use the simplest Abelian version of the SW model~\cite{son2} defined
by the 5D action
\begin{equation}
\label{1}
S=c^2\int d^4\!x\,dz\sqrt{g}\,e^{-az^2}\left(
-\frac{1}{4}F_{MN}F^{MN}+\frac12m_5^2V_MV^M\right),
\end{equation}
where $g=|\text{det}g_{MN}|$, $F_{MN}=\partial_M V_N-\partial_N
V_M$, $M,N=0,1,2,3,4$, $c$ is a normalization constant for the
vector field $V_M$, and the background space represents the Poincar\'{e} patch
of the AdS$_5$ space with the metric
\begin{equation}
\label{2}
g_{MN}dx^Mdx^N=\frac{R^2}{z^2}(\eta_{\mu\nu}dx^{\mu}dx^{\nu}-dz^2),\qquad z>0.
\end{equation}
Here $\eta_{\mu\nu}=\text{diag}(1,-1,-1,-1)$, $R$ denotes the radius of AdS$_5$ space,
and $z$ is the holographic coordinate which is usually interpreted as the inverse energy scale.
At each fixed $z$ one has the metric of flat 4D Minkowski space.
According to the standard prescriptions of AdS/CFT correspondence~\cite{witten,gub}
the 5D mass $m_5$ is determined by the behavior of $p$-form fields near the UV boundary $z=0$,
\begin{equation}
\label{3}
m_5^2R^2=(\Delta-p)(\Delta+p-4),
\end{equation}
where $\Delta$ means the scaling dimension of 4D operator dual to the corresponding 5D field on the UV boundary.
We consider the vector case, thus $p=1$ and $m_5^2R^2=(\Delta-1)(\Delta-3)$.
The minimal value of dimension for vector operator in QCD is $\Delta=3$ (the current~\eqref{h1}) that, according to~\eqref{3},
corresponds to massless 5D vector fields which are usually considered in the SW models.
But in general QCD operators interpolating vector mesons can have higher canonical dimensions, in particular,
as was discussed in section~2, the "descendants" preserving the chiral and Lorentz properties will
have dimensions
\begin{equation}
\label{3b}
\Delta=3+2k,\qquad k=0,1,2,\dots.
\end{equation}

The 4D mass spectrum of KK modes can be found, as usual, from the equation of
motion accepting the 4D plane-wave ansatz $V_M(x_\mu,z)=e^{ipx}v(z)\epsilon_\mu$
with the on-shell, $p^2=m^2$, and transverse, $p^\mu\epsilon_\mu=0$, conditions.
In addition, we will imply the condition $V_z=0$ for the physical
components of 5D fields~\cite{br3}. For massless vector fields, this is equivalent
to the standard choice of axial gauge due to emerging gauge invariance~\cite{son2}.
The ensuing from action~\eqref{1} equation of motion is
\begin{equation}
\label{4}
\partial_z\left(\frac{e^{-az^2}}{z}\partial_z v_n\right)=\left(\frac{m_5^2R^2}{z^3}-\frac{m_n^2}{z}\right)e^{-az^2}v_n.
\end{equation}
The particle-like excitations correspond to normalizable solutions of Sturm-Liouville equation~\eqref{4}.
It is known that they form an infinite discrete set $v_n(z)$.
The given property becomes more transparent after the substitution
\begin{equation}
\label{5}
v_n=z^{1/2}e^{az^2/2}\psi_n,
\end{equation}
which transforms the Eq.~\eqref{4} into a form of one-dimensional Schr\"{o}dinger equation
\begin{equation}
\label{6}
-\partial_z^2\psi_n+V(z)\psi_n=m_n^2\psi_n,
\end{equation}
with the potential
\begin{equation}
\label{7}
V(z)=a^2z^2+\frac{1+m_5^2R^2-1/4}{z^2},
\end{equation}
The mass spectrum of the model is given by the eigenvalues of Eq.~\eqref{6},
\begin{equation}
\label{8}
m_n^2=2|a|\left(2n+1+\sqrt{1+m_5^2R^2}\right),\qquad n=0,1,2,\dots.
\end{equation}
Using Eq.~\eqref{3} for $p=1$ the spectrum can be rewritten as
\begin{equation}
\label{9}
m_n^2=2|a|\left(2n+\Delta-1\right).
\end{equation}
The substitution of relation~\eqref{3b} into this spectrum leads to a remarkably simple prediction
\begin{equation}
\label{10}
m_n^2=4|a|\left(n+k+1\right),\qquad n,k=0,1,2,\dots.
\end{equation}
The formula~\eqref{10} demonstrates that within the standard SW holographic model,
the description of radial spectrum for vector mesons by interpolating HDO
is essentially the same as by KK modes --- the numbers $n$ and $k$ can be interchanged.

The spectral relation~\eqref{10} can be now used for relating the KK modes to excited mesons
in QCD. The lightest KK mode, $n=0$, corresponds to a vector state interpolated by the usual vector
current~\eqref{h1} of canonical dimension $\Delta=3$, the standard case of $\rho(770)$ and $\omega(782)$ mesons.
The mass of $n=1$ KK mode of field dual to the operator~\eqref{h1} coincides with the mass of $n=0$ mode
of field dual to the operator~\eqref{h4} with canonical dimension $\Delta=5$. We identify them and
interpret the given state as coupled to two interpolating operators. This can be continued
to arbitrary $n$-th radial excitation --- it will be coupled to a set of $n+1$ interpolating
operators with growing canonical dimensions. Thus the physical interpretation and description of radially
excited states become close in spirit to the description from first principles in lattice simulations~\cite{Dudek}.

It is not difficult to show that the same property and interpretation holds for other integer spins.
First let us consider the action for a free scalar field $\Phi$ on AdS$_5$ with the dilaton background as in~\eqref{1},
\begin{equation}
\label{11}
S=\frac12c^2\int d^4\!x\,dz\sqrt{g}\,e^{-az^2}\!\left(\partial_M\!\Phi\partial^M\!\Phi-m_5^2\Phi^2\right).
\end{equation}
As before, the physical 4D modes are given by the plane-wave ansatz $\Phi(x_\mu,z)=e^{ipx}\phi(z)$.
The equation of motion reads,
\begin{equation}
\label{12}
\partial_z\left(\frac{e^{-az^2}}{z^3}\partial_z \phi_n\right)=\left(\frac{m_5^2R^2}{z^5}-\frac{m_n^2}{z^3}\right)e^{-az^2}\phi_n,
\end{equation}
which after the substitution $\phi_n=z^{3/2}e^{az^2/2}\psi_n$
transforms into the Schr\"{o}dinger equation~\eqref{6}
with the potential
\begin{equation}
\label{14}
V(z)=a^2z^2+\frac{4+m_5^2R^2-1/4}{z^2}+2a.
\end{equation}
The corresponding eigenvalues are
\begin{equation}
\label{15}
m_n^2=2|a|\left(2n+1+\frac{a}{|a|}+\sqrt{4+m_5^2R^2}\right),\qquad n=0,1,2,\dots.
\end{equation}
Again making use of Eq.~\eqref{3} but for $p=0$ in~\eqref{15} we obtain
\begin{equation}
\label{16}
m_n^2=2|a|\left(2n+\frac{a}{|a|}+\Delta-1\right).
\end{equation}
Since the simplest scalar QCD operator shown in~\eqref{h2} has $\Delta=3$,
the relation~\eqref{3b} for dimensions of scalar "descendants" can be applied,
the radial scalar spectrum takes the form,
\begin{equation}
\label{17}
m_n^2=4|a|\left(n+k+1+\frac{a}{2|a|}\right),\qquad n,k=0,1,2,\dots,
\end{equation}
which differs from the vector spectrum~\eqref{10} only by a general shift
depending on the sign of constant $a$ in the dilaton background.

Now we will generalize the model to the case of arbitrary spin.
The higher spin fields $\Phi_J\doteq\Phi_{M_1M_2\dots M_J}$, $M_i=0,1,2,3,4$,
are described by symmetric, traceless tensors of rank $J$.
By assumption, the physical components of 5D fields satisfy the condition,
\begin{equation}
\label{18}
\Phi_{z\dots}=0,
\end{equation}
which is a generalization of condition $V_z=0$ used in the vector case.
The constraint $\partial^{\mu}\Phi_{\mu\dots}=0$ is also imposed to have
the required $2J+1$ physical degrees of freedom on the 4D boundary.
In general, an action for free higher spin fields contains many quadratic terms appearing from many ways of
contraction of coordinate indices. But the condition~\eqref{18} greatly simplifies the action leaving two terms only~\cite{son2,br3},
\begin{equation}
\label{19}
I=(-1)^{J}\frac12\int d^4x\,dz \sqrt{g}\,e^{-az^2}\!\left(\nabla_N\Phi_J\nabla^N\Phi^J-
m_J^2\Phi_J\Phi^J\right).
\end{equation}
It is known that the covariant derivatives $\nabla$ in the SW action~\eqref{19} can be effectively replaced by the usual ones
if the mass $m_J^2$ is shifted, $m_J^2\rightarrow b_Jz^2+c_J$, where $a_J$ and $b_J$ are certain $J$-dependent constants
emerging from affine connection in $\nabla$~\cite{br3}. They will entail a certain $J$-dependent contribution (a general shift)
to the spectrum of normalizable modes. The final effect of these terms turns out to be the same as if we used in~\eqref{19} just
normal derivatives and the five-dimensional mass were~\cite{br3,gutsche}
\begin{equation}
\label{3c}
m_J^2R^2=(\Delta-J)(\Delta+J-4).
\end{equation}
The $J$-dependent mass~\eqref{3c} is nothing but the relation~\eqref{3} with "$p$" replaced by "$J$".
As was demonstrated in Ref.~\cite{br2}, the origin of relation~\eqref{3c} lies in the
the condition~\eqref{18} which leads to decoupling of kinematical aspects from dynamical ones with final
outcome that symmetric tensors of rank $p$ (describing mesons of spin $p$) obey the same
equation as the $p$-form fields.

We will exploit all these findings to simplify the treatment of tensor mesons:
The covariant derivatives $\nabla$ in the action~\eqref{19} will be replaced by the normal ones
with simultaneous use of the relation~\eqref{3c}.
Imposing again the 4D plane-wave ansatz $\Phi_J(x_\mu,z)=e^{ipx}\phi^{(J)}(z)\epsilon_J$,
where $\epsilon_J$ denotes polarization,
we get the equation of motion for the profile function $\phi^{(J)}(z)$ of physical 4D modes,
\begin{equation}
\label{20}
-\partial_z\left(e^{-az^2}z^{2J-3}\partial_z\phi^{(J)}_n\right)+
m_J^2R^2e^{-az^2}z^{2J-5}\phi^{(J)}_n=m_n^2e^{-az^2}z^{2J-3}\phi^{(J)}_n.
\end{equation}
The substitution
$\phi^{(J)}_n=e^{az^2/2}z^{3/2-J}\psi^{(J)}_n$
casts the Eq.~\eqref{20} into the Schr\"{o}dinger equation~\eqref{6} with the potential
\begin{equation}
\label{21}
V(z)=a^2z^2+\frac{(J-2)^2+m_J^2R^2-1/4}{z^2}+2a(1-J).
\end{equation}
The discrete mass spectrum is
\begin{equation}
\label{22}
m_{n,J}^2=2|a|\left(2n+1+\frac{a}{|a|}(1-J)+\sqrt{(J-2)^2+m_J^2R^2}\right).
\end{equation}
Substituting relation~\eqref{3c}, we obtain the final result
\begin{equation}
\label{23}
m_{n,J}^2=2|a|\left(2n+\frac{a}{|a|}(1-J)+\Delta-1\right).
\end{equation}
It is easy to see that all corresponding equations for vector and scalar mesons considered above
represent just particular cases $J=1$ and $J=0$ of presented relations for arbitrary integer spin.
Now we need the last ingredient --- an extension of expression for canonical dimension of
interpolating operators~\eqref{3b} to the case of arbitrary spin. As follows from the discussions
in section~2, this extension of~\eqref{3b} is
\begin{equation}
\label{24}
\Delta=2+J+2k,\qquad k=0,1,2,\dots,
\end{equation}
which holds for $J\geq1$. From the Eq.~\eqref{23} we then obtain
for non-zero spins,
\begin{equation}
\label{25}
m_{n,J}^2=4|a|\left(n+k+\frac12(1+J)+\frac{a}{2|a|}(1-J)\right).
\end{equation}
For consistency, we must choose the sign $a<0$ in the dilaton background
because for $a>0$ the spectrum does not depend on spin, $m_{n,J}^2=4a(n+k+1)$.
Thus, the final relation is
\begin{equation}
\label{26}
m_{n,J}^2=4|a|(n+k+J),\qquad n,k=0,1,2,\dots,\quad J>0.
\end{equation}
The relation~\eqref{26} generalizes the spectrum of the simplest SW model~\cite{son2} for the twist-two interpolating operators ($k=0$)
to higher twists.
The choice of sign for $a$ is in fact predetermined by the way one introduces
the higher spin fields. If we introduced them following the original paper on SW model~\cite{son2}
as gauge massless fields on AdS$_5$, the situation would be opposite --- the mass spectrum~\eqref{26} would
be obtained for $a>0$ (this happens due to a specific rescaling of fields, see discussions on this point in Ref.~\cite{ahep}).
But independently of the issue of sign, we come to a remarkable conclusion that the integer number $n$ in~\eqref{26}
enumerating the KK excitation can be replaced by the integer number $k$ enumerating the higher-dimensional
interpolating operators with identical chiral and Lorentz properties which can be built in QCD.
Thus we obtain a new physical interpretation for this number
that historically gave rise to the notion of daughter Regge trajectories in old Veneziano dual amplitudes and in QCD string models.

It is important to emphasize that the structure of the Regge trajectories remains encoded in the geometry but in indirect way ---
via the boundary behavior of fields that dictates the $J$-dependence of 5D masses in the relation~\eqref{3c}.

We can get a further insight from consideration of normalized eigenfunctions corresponding to the discrete spectrum~\eqref{26},
\begin{equation}
\label{26b}
\phi^{(J)}_n=\sqrt{\frac{2n!}{(J+2k+n)!}}\,e^{-|a|z^2}\left(|a|z^2\right)^{1+k}L_n^{J+2k}\left(|a|z^2\right),
\end{equation}
where $L_n^\alpha(x)$ are associated Laguerre polynomials. It is seen that the numbers $n$ and $k$ are not completely interchangeable
in the radial wave function: While the large $z$ asymptotics depends on the sum $n+k$ (because $L_n^\alpha(x)\sim x^n$ at large $x$),
the number of zeros is controlled by $n$ only (as the polynomial $L_n^\alpha(x)$ has $n$ zeros). By setting $n=0$, i.e. by keeping
the zero KK mode only, we thus choose the wave function without zeros in holographic coordinate. This wave function is the least
"entangled" with the 5th holographic dimension and thereby is the least sensitive to deviations from the AdS structure. As we will
discuss in section~5, this makes the zero KK mode the most reliable in the phenomenological holographic approaches.

The real QCD operators have anomalous dimensions and this represents a notorious problem for the whole bottom-up holographic approach.
One makes reference to asymptotic freedom at best, any serious discussion of this problem is usually avoided. We will not give
a real physical justification but make an observation. Within our considerations, the account for the anomalous dimension of
operators is tantamount to replacement $k\rightarrow k+\varepsilon(k,J)$ in the spectrum~\eqref{26}. Then $2\varepsilon(k,J)$
(see Eq.~\eqref{3b}) reflects contribution to the canonical dimension $\Delta$ from the anomalous part. The systematic form of
$\varepsilon(k,J)$ is unknown but it is naturally expected that $\varepsilon(k,J)$ is a growing function of both arguments.
However, the spectrum~\eqref{26} more or less meets the existing phenomenology~\cite{shif,deg,bugg}. This should mean that
$\varepsilon(k,J)$ is either suppressed in the large-$N_c$ limit (perhaps the size of $\varepsilon(k,J)$ can be then systematically
estimated by a phenomenological analysis of deviations from the relation~\eqref{26}) or by itself is an approximate linear function
of its arguments (hence, the effects of anomalous dimensions are then effectively absorbed by the phenomenological values of
parameters in~\eqref{26}). The both possibilities could constitute an interesting prediction of the SW holographic approach.

\section{The Hard Wall model}

The spectrum of HW holographic model~\cite{son1,pom} is not Regge like, on the other hand, this model is much better
accommodated for description of the chiral symmetry breaking in QCD and hadron form factors. The analysis of previous section can be
easily applied to the HW model --- one can just set $a=0$ in the corresponding equations and impose the infrared
cutoff $z_m$. The normalizable solution of Eq.~\eqref{20} satisfying $\phi(0)=0$ is then given by
\begin{equation}
\label{27}
\phi\sim z^{2-J}J_{\Delta-2}(mz),
\end{equation}
where $J_\alpha(x)$ is the Bessel function. The solution~\eqref{27} corresponds to interpolating QCD operator
of canonical dimension $\Delta$. In the original papers~\cite{son1,pom}, the discrete spectrum
emerges from the Dirichlet boundary condition,
\begin{equation}
\label{28}
\partial_z\phi(m_nz_m)=0.
\end{equation}

We find more convenient to introduce the higher spin fields into the HW model in the gauge invariant way ---
the way adopted in the original SW model~\cite{son2}.
Actually this analysis in the HW model was carried out for fields
dual to twist-2 operators in Ref.~\cite{katz}, we can directly take the final result and generalize it to arbitrary twists.
In brief, the ensuing
equation of motion is tantamount to making the substitution $\phi^{(J)}=z^{4-J-\Delta}\tilde{\phi}^{(J)}$ (the field
$\tilde{\phi}^{(J)}$ becomes constant at the UV boundary $z\rightarrow0$) and dropping the 5D mass term in Eq.~\eqref{20}
with $a=0$. We get
\begin{equation}
\label{32}
-\partial_z\left(z^{1-2(\Delta-2)}\partial_z\tilde{\phi}^{(J)}_n\right)=m_n^2z^{1-2(\Delta-2)}\tilde{\phi}^{(J)}_n.
\end{equation}
The normalizable solution satisfying $\tilde{\phi}^{(J)}(0)=0$ is
\begin{equation}
\label{33}
\tilde{\phi}^{(J)}\sim z^{\Delta-2}J_{\Delta-2}(mz).
\end{equation}
The Dirichlet boundary condition leads to the following equation for discrete spectrum,
\begin{equation}
\label{34}
J_{\Delta-3}(m_nz_m)=0,
\end{equation}
where the property of Bessel functions $\partial_x \left(x^\alpha J_\alpha\right)=x^\alpha J_{\alpha-1}$
was exploited. Setting $\Delta=J+2$ in Eq.~\eqref{34}, we obtain the equation of Ref.~\cite{katz} for spectrum
of higher spin mesons interpolated by twist-2 operators.

Substituting to the Eq.~\eqref{34} different canonical dimensions $\Delta$ we will get
radial spectra which do not coincide in any part because zeros of different Bessel functions
have different set of locations. This shows that the KK modes of HW model can not be traded for
states arising from HDO.

It is interesting to notice, however, that if we consider the spectrum of $n=0$ modes for $\Delta=3,4,5,\dots$
in the Eq.~\eqref{34}, we get the sequence of roots: $m_0(\Delta)z_m\approx\{2.4,3.8,5.1,6.4,7.6,\dots\}$.
For the usual inputs $z_m^{-1}\approx323$~MeV and $m_0(\Delta=3)=776$~MeV of the HW model~\cite{son1},
these roots lead to the mass spectrum (in MeV)
\begin{equation}
\label{35}
m_0(\Delta)\approx\{776,1234,1653,2056,2452,\dots\}.
\end{equation}
The given spectrum yields much
slower rise of masses in the radial trajectories than in the traditional HW approach (namely $m_{\rho,n}\approx\{776,1777,2810,3811,\dots\}$ MeV
in Ref.~\cite{son1}). This provides a qualitatively better spectroscopy: In the mass range below 2.5~GeV
(relatively well scanned experimentally for light non-strange mesons) we obtain 5 $\rho$-mesons that is more or less in accord with the Particle Data~\cite{pdg},
while the standard HW model predicts only 2 $\rho$-mesons.
It is also curious to observe that the spectrum~\eqref{35} interpolates with a good precision the experimental positions for clusters of light
non-strange meson resonances arising from approximate mass degeneracies between radial and spin (or orbital) excitations~\cite{shif,deg,bugg}.
This may be viewed as a phenomenological indication that in the HW model, only zero $KK$ modes of fields dual to HDO in QCD have
physical significance.

\section{Discussions}

The KK modes are usually regarded as inherent for holographic models and are widely used in phenomenological
applications. As we mentioned in Introduction, their physical properties, however, are very different from those of
hadron resonances~\cite{csaki}. Below we draw attention to the fact that the KK modes and resonances in confining
theory, in a sense, come from different sectors from the point of view of original
AdS/CFT correspondence~\cite{mald}.

In the compactified AdS space of HW models, the nature of KK modes is similar to KK modes from flat
extra space compactified on a circle --- they emerge from boundary conditions on propagation of particles in the {\it whole}
extra dimension and the higher is a KK-mode, the stronger is "entangled" its wave function with extra dimension (has more zeros).
In the infinite AdS space --- the case of SW model --- the nature of KK modes becomes more tricky.
The gravitational potential grows in spaces with negative lambda-term, as a result the gravity acts
on massive particles as a kind of potential well~\cite{zwiebach}. Massless particles in AdS are known to reach
the boundary at infinity for finite time~\cite{zwiebach}, so they can be also considered as living in a gravitational
box. As before, the KK modes are well-defined if the {\it whole} infinite AdS space is available for propagation.
The problem of reliability of KK modes in practical holographic models arises from the fact that the AdS$_5$ space in the
AdS/CFT correspondence is an approximation which is valid only in the deep infrared domain.

To clarify the point, it should be recalled that the Maldacena's gauge/gravity conjecture originated
from the three-brane supergravity solution for a stack of $N_c$ coincident 3D branes in 10D ambient space~\cite{mald}.
The corresponding extremal solution for the sign convention as in~\eqref{2} has the form (omitting the
4-form RR-potential)
\begin{equation}
\label{d1}
ds^2=\left(1+\frac{R^4}{r^4}\right)^{-\frac12}dy_\mu^2-\left(1+\frac{R^4}{r^4}\right)^{\frac12}\left(dr^2+r^2d\Omega_5^2\right),
\end{equation}
$$
r^2=\sum_{i=4}^{9}y_i^2,\qquad R^4=4\pi g_s N_c l_s^4=g_\text{YM}^2N_c l_s^4=\lambda l_s^4,
$$
where the indices $\mu=0,1,2,3$ run along the worldvolume of D3 branes, $\Omega_5$ denotes unit 5-sphere,
$l_s$ and $g_s$ are the string length and coupling. The gravitational description is justified for
$R\gg l_s$ that entails the large 't Hooft coupling $\lambda \gg 1$. The conjecture of AdS/CFT correspondence
emerges in the region $r\ll R$, where the metric~\eqref{d1} factorizes into AdS$_5\times$S$_5$,
\begin{equation}
\label{d2}
\left.ds^2\right|_{r\ll R}=\frac{r^2}{R^2}dy_\mu^2-\frac{R^2}{r^2}dr^2-R^2d\Omega_5^2.
\end{equation}
Strictly speaking, one obtains in~\eqref{d2} the Poincar\'{e} patch of AdS$_5$ describing only half of the global AdS$_5$ space
(a region $r>0$ lying from "one side" of the D3 stack). After the change of coordinate $r=R^2/z$ this patch takes
the form of~\eqref{2}. We see thus that the metric~\eqref{d2} in holographic QCD has a justification
within the original gauge/gravity duality for $z\gg R$ only. The matching to QCD, however, is usually performed
in the limit $z\rightarrow 0$ where the metric~\eqref{d1} does not have an AdS part at all.
Since the original AdS/CFT correspondence does not have matter in the fundamental
representation, a better example could be the top-down holographic approach~\cite{erdmenger} --- after geometrical introduction
of flavor branes and of chiral symmetry breaking, the metric of resulting low energy models has nothing in common with AdS.

These remarks may suggest that the higher KK modes lie outside the applicability
of the bottom-up holographic approach. On the other hand, they are indispensable for internal consistency as
they provide the correct high energy behavior of correlation functions. According to the analysis of our work,
these higher KK modes can be effectively replaced by infinite number of zero KK modes stemming from heavier fields
dual to HDO in QCD. This removes the aforementioned discrepancy.
Perhaps one can develop a more solid justification by imposing the
Pauli-Villars regularization in line with the compactified AdS$_5$ (i.e. Randall-Sundrum) extra-dimensional
scenario of Ref.~\cite{Pomarol:2000hp} --- this regularization effectively suppresses the propagation of KK modes of gauge fields
except zero mode.



The linear form of radial spectrum of SW model in our scheme remains unchanged and the proposed approach by itself
cannot explain the origin of quadratic dilaton background in the action~\eqref{1}. This long-standing problem is still open.
The original proposal for the dilaton factor $e^\varphi$ was inspired by the string theory but the quadratic form,
$\varphi=-az^2$, speculatively conceivable as a possible result of "closed string tachyon condensation"~\cite{son2},
was completely phenomenological. A gravitational theory with quadratic dilaton background is not known. There are
numerous attempts in the literature to develop a sort of dynamical Maxwell-dilaton-gravity model, however, the
desired background $e^{-az^2}$ on AdS$_5$ space either does not appear even approximately or is obtained at the cost
of introduction of some complicated potential $V(\varphi)$ in the action. Without any theoretical restrictions on
the form of $V(\varphi)$ (e.g., like renormalizability) such a solution looks as {\it ad hoc} as introduction of this
background  by hands. Some relevant discussions of this problem are given in Ref.~\cite{sonn}

It is not excluded that a viable solution lies in a completely different direction. For instance, one may speculate
that this background can be interpreted as an effective way for taking into account some quantum features of
particle propagation in the AdS space-time, in which gravity effectively acts as a finite box. In this respect,
it is interesting to mention the recent work~\cite{Fichet:2019hkg} where it was demonstrated that in Lorentzian
AdS, gravity dresses free propagators on the quantum level. In particular, the scalar component of 5D graviton
leads to universal exponential suppression of propagators in the infrared region,
\begin{equation}
\label{d3}
\Delta_p(z)\propto e^{-\alpha pz},
\end{equation}
where $p=\sqrt{p^\mu p_\mu}$ and positive constant $\alpha\sim 1/(RM_\text{Pl})^2$ arises from one-loop gravitational
corrections. The one-loop corrections due to interactions with other fields in the bulk (including self-interaction)
also lead to the suppression of the kind~\eqref{d3}~\cite{Fichet:2019hkg}.
In other words, the AdS space turns out to be opaque to propagation in deep infrared region.
Basing on this result, we may imagine the following. At high enough energy one has $p\sim E_\text{p}$, where
$E_\text{p}$ is the particle proper energy (the energy measured in the reference frame instantaneously at rest at $z$).
The proper energy is related to the energy in the gauge theory $E_\text{exp}$, that is measurable experimentally,
via the rescaling dictated by the AdS metric~\eqref{2},
\begin{equation}
\label{d4}
p\sim E_\text{p}=\frac{z}{R}E_\text{exp}.
\end{equation}
Then if we want to deal with measurable quantities (as we do in the bottom-up holographic QCD)
we should substitute~\eqref{d4} into~\eqref{d3},
\begin{equation}
\label{d5}
\Delta_p(z)\propto e^{-|a|z^2},
\end{equation}
where $a$ is a constant proportional to a typical energy scale multiplied by AdS$_5$ curvature $1/R$ and by
5D gravity coupling squared from one-loop gravitational corrections. Loosely speaking, the dim2 slope parameter $a$ of
meson trajectories emerges as a result of multiplication of two dim1 factors stemming from the UV and IR dynamics
correspondingly. A similar interpretation for $a$ was proposed in Refs.~\cite{plb} from different considerations.

The asymptotic behavior~\eqref{d5} can be compared with the bulk-to-boundary propagators in the SW model.
In the vector case~\eqref{1} (see, e.g., Ref.~\cite{Afonin:2012xq}),
\begin{equation}
\label{d6}
\Delta_p(z)=\Gamma\left(1-\frac{p^2}{4|a|}\right)e^{(a-|a|)z^2/2}U\left(\frac{-p^2}{4|a|},0;|a|z^2\right),
\end{equation}
where $U(\alpha,\beta;x)$ is the Tricomi confluent hypergeometric function which vanishes for
$x\rightarrow\infty$ as $x^{-\alpha}$. For arbitrary spin $\Delta_p(z)$ is qualitatively similar ---
the main change is $\beta=|J-1|$ in $U(\alpha,\beta;x)$~\cite{br3}. Now one can observe that at $a<0$
in~\eqref{d6} --- the choice that we made in section~3 for arbitrary spin --- the propagators in the SW model
acquire the same exponential suppression as in~\eqref{d5}. The dilaton background of the SW model might be
thus viewed as an effective account for corrections to particle propagators in the AdS$_5$ space stemming from
quantum gravity and other possible interactions in the bulk.

\section{Conclusions}

In the usual AdS/QCD approach, the radially excited hadrons are identified with Kaluza-Klein excitations
of dual theory. This has, however, severe theoretical problems. We proposed an alternative description in
which the  excited modes become a superposition of several Kaluza-Klein states arising from massive 5D fields
dual to QCD operators with different canonical dimensions.
Exactly along this line the radially excited states are extracted in the lattice simulations from a set
of interpolating QCD operators.
The new description was applied to the Soft Wall
and Hard Wall holographic models in the sector of light mesons. The linear radial and
angular Regge spectrum of Soft Wall model remains unchanged. The radial spectrum of Hard Wall model changes
significantly. We consider this result as another strong indication (in addition to non-Regge behavior) on the fact
that the KK modes of HW model can not be identified with real mesons in the large-$N_c$ limit of QCD.

The approach proposed in the present work may be interpreted as a simple tool establishing a direct connection between
composite QCD operators and hadron resonances with a definite mass.
The emerging connection becomes thus a new nice feature of bottom-up holographic models.
Contrary to many theoretical approaches aimed at description of excited hadrons,
the description in AdS/QCD becomes close in spirit to the methods used in lattice simulations~\cite{Dudek}.

As the presented holographic approach to hadron spectroscopy is more tightly related with the real QCD than the traditional
bottom-up holographic models, it would be interesting to explore it beyond the hadron spectroscopy. We leave this
work for future.


\begin{thebibliography}{99}

\bibitem{hoof} G.~'t Hooft, Nucl. Phys. B~{\bf 72}, 461 (1974).
\bibitem{wit}  E.~Witten, Nucl. Phys. B~{\bf 160}, 57 (1979).
\bibitem{mald} J.~M.~Maldacena, Adv. Theor. Math. Phys. {\bf 2}, 231 (1998);
Int. J. Theor. Phys. {\bf 38}, 1113 (1999).
\bibitem{witten} E. Witten, Adv. Theor. Math. Phys. {\bf 2},  253 (1998).
\bibitem{gub} S.~S.~Gubser, I.~R.~Klebanov and A.~M.~Polyakov, Phys. Lett. B {\bf 428}, 105 (1998).
\bibitem{son1} J.~Erlich, E.~Katz, D.~T.~Son and M.~A.~Stephanov,
Phys. Rev. Lett. {\bf 95}, 261602 (2005).
\bibitem{pom} L.~Da Rold and A.~Pomarol,
Nucl. Phys. B {\bf 721}, 79 (2005).
\bibitem{son2} A.~Karch, E.~Katz, D.~T.~Son and M.~A.~Stephanov,
Phys. Rev. D {\bf 74}, 015005 (2006).
\bibitem{csaki} C.~Csaki, M.~Reece and J.~Terning,
JHEP {\bf 0905}, 067 (2009).
\bibitem{Dudek}
  J.~J.~Dudek {\it et al.},
  Phys.\ Rev.\ D {\bf 82}, 034508 (2010).
\bibitem{MMM} G.~Colangelo, F.~Hagelstein, M.~Hoferichter, L.~Laub and P.~Stoffer,
JHEP \textbf{03}, 101 (2020);
J.~Leutgeb and A.~Rebhan,
Phys. Rev. D \textbf{101}, 114015 (2020);
L.~Cappiello, O.~Cata, G.~D'Ambrosio, D.~Greynat and A.~Iyer,
arXiv:1912.02779 [hep-ph].
\bibitem{qqm}
  A.~A.~Andrianov and V.~A.~Andrianov,
  Int.\ J.\ Mod.\ Phys.\ A {\bf 8}, 1981 (1993);
  A.~A.~Andrianov, V.~A.~Andrianov and V.~L.~Yudichev,
  Theor.\ Math.\ Phys.\  {\bf 108}, 1069 (1996)
  [Teor.\ Mat.\ Fiz.\  {\bf 108}, 276 (1996)];
  V.~A.~Andrianov and S.~S.~Afonin,
  Eur.\ Phys.\ J.\ A {\bf 17}, 111 (2003);
  Zap.\ Nauchn.\ Semin.\  {\bf 291}, 5 (2002)
  [hep-ph/0304140].
\bibitem{chizh}
  M.~V.~Chizhov,
  hep-ph/9610220.
\bibitem{multiplets}
  T.~D.~Cohen and X.~D.~Ji,
  Phys.\ Rev.\ D {\bf 55}, 6870 (1997);
  M.~Shifman,
  hep-ph/0507246;
  L.~Y.~Glozman,
  Phys.\ Rept.\  {\bf 444}, 1 (2007).
\bibitem{shif} M. Shifman and A. Vainshtein, Phys. Rev. D {\bf 77}, 034002 (2008).
\bibitem{pdg} M. Tanabashi {\it et al.} (Particle Data Group), Phys. Rev. D {\bf 98}, 030001 (2018).
\bibitem{svz} M. A.~Shifman, A.~I.~Vainstein and V.~I.~Zakharov, Nucl. Phys.
B~{\bf 147}, 385, 448 (1979);
L.J. Reinders, H. Rubinstein, S. Yazaki, Phys. Rept. {\bf 127}, 1 (1985);
P.~Colangelo and A.~Khodjamirian,
  hep-ph/0010175.
\bibitem{deg} S.~S. Afonin,  Eur. Phys. J. A {\bf 29}, 327 (2006);
Phys. Lett. B {\bf 639}, 258 (2006); Phys. Rev. C {\bf 76}, 015202 (2007);
Mod. Phys. Lett. A {\bf 22}, 1359 (2007); Int. J. Mod. Phys. A {\bf 22}, 4537 (2007);
E. Klempt and A. Zaitsev, Phys. Rept. {\bf 454}, 1 (2007).
\bibitem{br3}
  S.~J.~Brodsky, G.~F.~de Teramond, H.~G.~Dosch and J.~Erlich,
  Phys.\ Rept.\  {\bf 584}, 1 (2015).
\bibitem{gutsche} T. Gutsche, V. E. Lyubovitskij, I. Schmidt and A. Vega,
Phys. Rev. D {\bf 85}, 076003 (2012).
\bibitem{br2} G.~F.~de Teramond and S.~J.~Brodsky, Phys. Rev. Lett. {\bf 102}, 081601 (2009);
G. F. de Teramond, H. G. Dosch and S. J. Brodsky,
Phys. Rev. D {\bf 87}, 075005 (2013).
\bibitem{ahep}
  S.~S.~Afonin,
  Adv.\ High Energy Phys.\  {\bf 2017}, 8358473 (2017).
\bibitem{katz} E.~Katz, A.~Lewandowski and M.~D.~Schwartz,
Phys. Rev. D {\bf 74}, 086004 (2006).
\bibitem{bugg} D.~V.~Bugg, Phys. Rept. {\bf 397}, 257 (2004).
\bibitem{zwiebach}B. Zwiebach, {\it A First Course in String Theory}, Cambridge University Press; 2nd edition.
\bibitem{erdmenger} J.~Erdmenger, N.~Evans, I.~Kirsch and Threlfall,
Eur. Phys. J. A \textbf{35}, 81 (2008).
\bibitem{Pomarol:2000hp}
  A.~Pomarol,
  Phys.\ Rev.\ Lett.\  {\bf 85}, 4004 (2000).
\bibitem{sonn} J.~Sonnenschein,
Prog. Part. Nucl. Phys. \textbf{92}, 1 (2017).
\bibitem{Fichet:2019hkg}
  S.~Fichet,
  Phys.\ Rev.\ D {\bf 100}, 095002 (2019).
\bibitem{plb} S.~S.~Afonin,
  Phys.\ Lett.\ B {\bf 675}, 54 (2009);
  Phys.\ Lett.\ B {\bf 678}, 477 (2009).
\bibitem{Afonin:2012xq}
  S.~S.~Afonin,
  Int.\ J.\ Mod.\ Phys.\ A {\bf 27}, 1250171 (2012).










\end{thebibliography}
\end{document}